# ИНФОРМАТИКА

УДК 004.9:66.013.512

## МЕТОДЫ ОБЕСПЕЧЕНИЯ ТРЕБОВАНИЙ ЕСКД И СПДС ПРИ РАЗРАБОТКЕ САПР ПРОМЫШЛЕННЫХ ОБЪЕКТОВ

В.В.Мигунов

*ЦЭСИ РТ при КМ РТ, г. Казань*

Аннотация

Изложены методы обеспечения требований российских стандартов в САПР промышленных объектов, реализованные в отечественной системе TechnoCAD GlassX, имеющей собственное графическое ядро и собственные структуры хранения данных. Показано, что привязка структур хранения и программного кода САПР к требованиям стандартов дает возможность не только выполнить эти требования в проектной документации, но и повысить степень компактности хранения чертежей как на диске, так и в оперативной памяти.

Библ.7

Abstract

V.V. Migunov. The methods of support of the requirements of the Russian standards at development of a CAD of industrial objects // Izvestiya of the Tula State University/ Ser. Mathematics. Mechanics. Informatics. Tula: TSU, 2004. V._. N _. P. __–__.

The methods of support of the requirements of the Russian standards in a CAD of industrial objects are explained, which were implemented in the CAD system TechnoCAD GlassX with an own graphics core and own structures of data storage. It is rotined, that the binding of storage structures and program code of a CAD to the requirements of standards enable not only to fulfil these requirements in project documentation, but also to increase a degree of compactness of storage of drawings both on the disk and in the RAM.

Bibl.7

*Введение*

Широко известна проблема соблюдения требований российских стандартов единой системы конструкторской документации (ЕСКД) и системы проектной документации для строительства (СПДС) при разработке САПР на основе иностранных универсальных графических ядер. Для подготовки чертежей СПДС не требуются многие их возможности, отвечающие задачам параметрического трехмерного представления и глубоко разработанные в связи с потребностями машиностроения. Также на практике ограничивается круг используемых типов и толщин линий, цветов [1]. В то же время имеются затруднения при выполнении чертежей, например, средствами AutoCAD™ во фронтальной диметрической проекции – это не изометрическая проекция. В результате разрабатываются специализированные надстройки, такие, как auto.ЕСКД и auto.СПДС [2]. Начавшие появляться в этом десятилетии комплексные САПР промышленных объектов [3, 4] в первую очередь характеризуются их разработчиками как отвечающие ЕСКД и СПДС, как русифицированные и имеющие информационное обеспечение в соответствии с российскими стандартами. В [3] говорится, что "требования к проектированию с использованием российской базы данных и по российским стандартам можно отнести к обязательным задачам, решаемым современной средой проектирования", что представляется вполне справедливым.

В случае, когда САПР создается на основе собственного графического ядра, появляется возможность снять эту проблему естественным путем, без специальных надстроек. С точки зрения программной реализации здесь можно выделить два уровня отсечения вариантов, не соответствующих стандартам. Первый, менее жесткий уровень, ограничивается запретами на стадии ввода данных, он реализуется в пользовательском интерфейсе. При этом в структурах хранения данных остается универсальность, как в иностранных графических ядрах. Второй, жесткий уровень - когда сами структуры хранения не позволяют внести в электронный чертеж информацию, противоречащую стандартам. На втором уровне дополнительно достигается сокращение размеров чертежей в оперативной памяти и на диске, и, соответственно, повышение быстродействия САПР. В любом случае вопрос учета требований стандартов должен рассматриваться при разработке как структур хранения данных, так и программного кода САПР.

Произошедший за последние пятнадцать лет переход к открытой экономике привел к тому, что задачи строительства и реконструкции предприятий часто решаются с привлечением иностранных технологий и оборудования. При этом проекты включают части, разработанные иностранными поставщиками по их стандартам. Например, в технологических регламентах химических производств неотъемлемой частью является технологическая схема, на чертеж которой наносятся условные графические обозначения приборов и исполнительных механизмов, несколько отличающиеся от отечественных. По этим причинам, а также в связи с процессами изменения действующих стандартов и норм, наличием в них правил рекомендательного характера, оптимальным представляется компромиссный вариант, когда в структурах хранения предусматривается некоторая избыточность, а актуальные ограничения и их изменения отражаются прежде всего в программном коде, где происходит интерпретация данных.

В литературе отсутствует подробный теоретический анализ методов параметризации, применяемых в популярных САПР [5], не распространяются и сведения о внутренних форматах хранения данных. По существу, структуры хранения определяют параметрические представления объектов и являются технологическим ноу-хау разработчиков, которые не заинтересованы в их раскрытии по причинам коммерческого характера [5]. Сложившаяся ситуация замедляет нразвитие САПР. Настоящая работа посвящена методам обеспечения требований ЕСКД и СПДС, которые реализованы в отечественной САПР реконструкции предприятий TechnoCAD GlassX [6] на основе собственного графического ядра. Приводимые сведения могут быть полезны при разработке САПР промышленных объектов.

*Общие методы*

Чертеж представляется как совокупность геометрических элементов, каждый из которых изображается одним из 6 типов линий, разрешенных ГОСТ 2.303, с точки зрения толщины и продольной структуры. Шаг разрывов в штриховых и штрих-пунктирных линиях фиксируется как средний из допускаемых стандартом. Кроме этого, добавлен допускаемый по ГОСТ 21.106 тип линии утолщенная штриховая. Идентификация типа линий занимает 3 бита. Под цвет отводится 4 бита, и один бит - под признак принадлежности системе координат Натура или Бумага (в чертеже имеется две системы координат с общим началом и направлениями осей, отличающиеся масштабом).

Масштаб чертежа выбирается только из вариантов по ГОСТ 2.302, причем в специализированных проблемно-ориентированных расширениях варианты масштабов ограничиваются в соответствии с требованиями СПДС для каждого конкретного случая. Для профилей наружных сетей водоснабжения и канализации, например, по горизонтали допускаются масштабы от 1:500 до 1:5000, а по вертикали - от 1:100 до 1:500 (ГОСТ 21.604).

Чертежный шрифт допускается по ГОСТ 2.304 с вариантами размеров 2.5, 3.5, 5,7, 10, 14, 20, 28, 40 и углов наклона 90° либо 75°, заданных в стандарте - всего 18 вариантов, хранимых в 8 битах (7 из них заняты под размер шрифта, хранятся варианты с шагом 0,5 мм), рис.1.

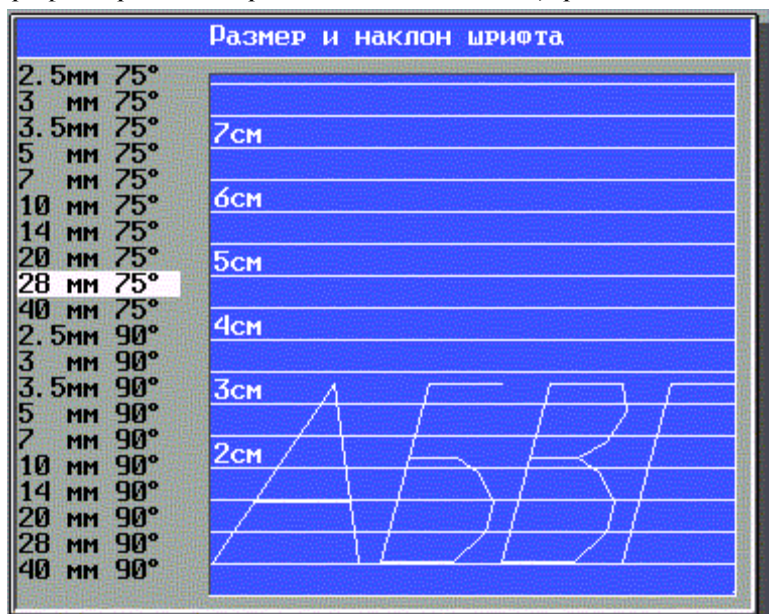

Рис.1. Выбор в меню допустимых по ЕСКД размеров и углов наклона шрифта

При нанесении размеров величина стрелок, засечек, перехода выносных линий за размерные и др. могут выбираться только в диапазоне, задаваемом ГОСТ 2.307. Структуры хранения расширяют диапазон в 2-3 раза, дискретность хранения 0,2мм.

Формат выбирается из вариантов ГОСТ 2.301, включая дополнительные форматы. Пользовательские форматы разрешены, но их размер надо задавать вручную, что затрудняет работу.

Дискретность - 1 мм.

Режим ортогональности при черчении и режим подготовки аксонометрических схем предусматривает координатные оси в 13 допустимых по ГОСТ 2.317 аксонометрических проекциях, а также в 6 более простых видах по ГОСТ 2.305, плюс дополнительные 6 косоугольных фронтальных проекций (3 изометрические и 3 диметрические) с направлением оси X горизонтально вправо и направлением оси Y в первом квадранте, возникшим из примеров чертежей в СПДС. Проекции идентифицируются номером. В специализированном режиме подготовки аксонометрических схем трубопроводных систем допускаются также произвольные изометрические проекции, которые сохраняются в параметрическом представлении схем специальным образом.

Для специфицирования чертежа в него помещаются таблицы, выбираемые из 22 вариантов по ГОСТ 2.108, 21.103, 21.401, 21.101, 21.110, по типовым проектным решениям и стандартам предприятия. Кроме этого, имеются еще несколько бланков таблиц специального вида: ведомость перемычек, потребность кабелей и проводов и др. по СПДС. Возможно создание новых бланков таблиц, но удобнее использовать имеющиеся.

Библиотеки условных графических обозначений, помещаемых в чертеж, включают только заданные в ГОСТ варианты. То же касается библиотеки основных и дополнительных надписей, рабочих чертежей элементов трубопроводов, элементов сварных швов и других служебных библиотек, которые пользователь не может изменить.

Способы штрихования предусматривают набор из ГОСТ 2.306 и другие стили, применяемые в чертежах марок АР, КЖ, КМ СПДС. Можно создать и свой стиль штрихования, но это труднее, чем выбрать имеющийся.

При специфицировании изделий, изображенных на чертеже, осуществляется выбор в электронных номенклатурных каталогах, которые полностью основаны на стандартах, технических условиях и информации заводов-изготовителей, причем автоматическое формирование обозначений и наименований изделий происходит в точности по тем правилам, которые заданы в исходной нормативной документации.

*Методы, основанные на модульной технологии*

Следующие далее методы обеспечения требований стандартов реализованы за счет автоматической генерации изображений по их параметрическому представлению. Используется технология, основанная на совместном хранении в одном элементе чертежа, называемом "Модуль", как исходных данных, так и результатов параметрической генерации. Модуль включает видимую в чертеже совокупность геометрических элементов и невидимое в чертеже параметрическое представление моделируемого объекта. При внесении изменений в параметры геометрическая часть модуля генерируется заново.

Условные обозначения приборов и исполнительных механизмов в схемах автоматизации генерируются по буквенно-цифровым признакам в точном соответствии с требованиями ГОСТ 21.404. Таблица расположения средств автоматизации (ГОСТ 21.408) автоматически генерируется с необходимыми размерами граф, типами линий и шрифтами, задаются и хранятся только размеры полей.

При подготовке специфицирующих таблиц их размеры автоматически соответствуют стандартным, то же для используемых типов линий, правил продолжения таблиц, шрифтов в них. Автоматизируется простановка и упорядочивание позиций по ГОСТ 2.108 и специализированным ГОСТам СПДС. Предусматривается возможность автоматической конвертации заказной спецификации из формата ГОСТ 21.103-78 в формат ГОСТ 21.110-95.

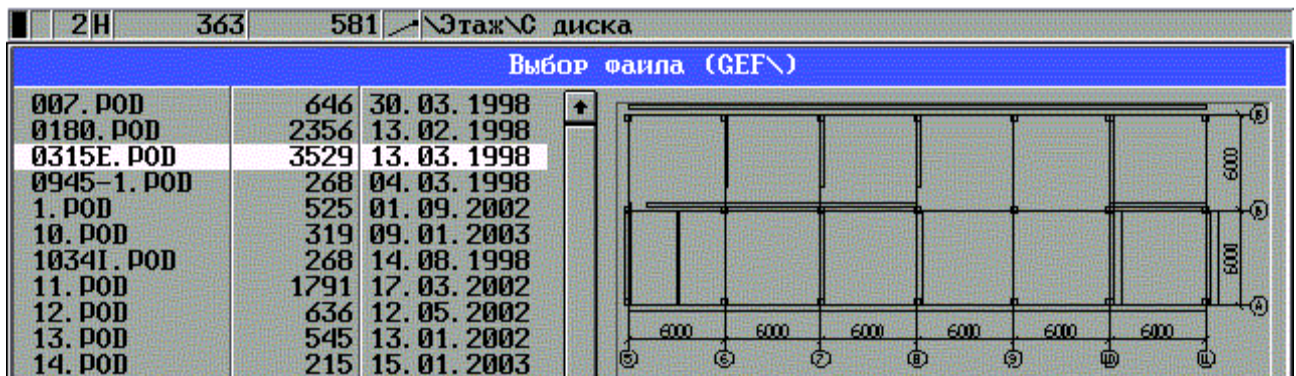

Рис.2. Выбор прототипа из дисковых комплектов параметров строительной подосновы

В модулях поэтажных планов (рис.2) строительной подосновы обозначения осей, размеры и многое другое генерируется в соответствии с ГОСТ 21.501 по задаваемым количествам вертикальных и горизонтальных осей. Автоматически наносятся перегородки выбранного типа, дверные и оконные проемы и др. Компактность параметрических представлений в модульной технологии характеризуется, например, тем, что комплект параметров плана этажа на рис.2 занимает на диске 3529 байт.

Аксонометрические схемы трубопроводов используют стандартную библиотеку условных графических обозначений трубопроводной арматуры и элементов трубопроводов. Сами схемы, включая нанесение текстов и размеров, обозначений разрывов, изображаются по ГОСТ 21.206, 21.401, 21.601, 21.602. При нанесении условных графических обозначений плоскость обозначения автоматически подбирается так, чтобы выносные линии лежали одновременно в этой плоскости и в одной из координатных плоскостей (не более 6 вариантов). По-другому разместить обозначение нельзя. Автоматизировано нанесение разрывов на трубы.

Геометрическая часть модуля профиля наружной сети водоснабжения или канализации генерируется по ГОСТ 21.604, включая условные обозначения труб, футляров, колодцев, дождеприемников, таблицу основных данных, нанесение текстов, размеров и отметок высоты.

Модуль с проектом молниезащиты включает в соответствии с требованиями [7] обязательный вид сверху (план) и необязательные дополнительные сечения в вертикальных плоскостях для систем молниезащиты из стержневых (одиночных, двойных и многократных), тросовых (одиночных и двойных) и сеточных молниеприемников. Расчет сечений зон защиты и их взаимовлияния ведется автоматически по правилам [7], также как и генерация геометрии, включающей: условные изображения стержневых, сеточных, одиночных и двойных тросовых молниеприемников, позиционные обозначения, обозначение вида сверху, обозначения вертикальных сечений на виде сверху (включая изображение секущего отрезка) и на самом вертикальном сечении, размеры, линии горизонтальных и вертикальных сечений зон защиты, порождаемых различными наборами молниеприемников, текстовые обозначения сечений на виде сверху с указанием высот, таблица расчета молниезащиты, изображения заземлителей.

Оформление чертежа рамками, основной и дополнительными надписями вычерчивается по требованиям стандартов автоматически из параметрического представления в модуле оформления чертежа.

*Магистрали*

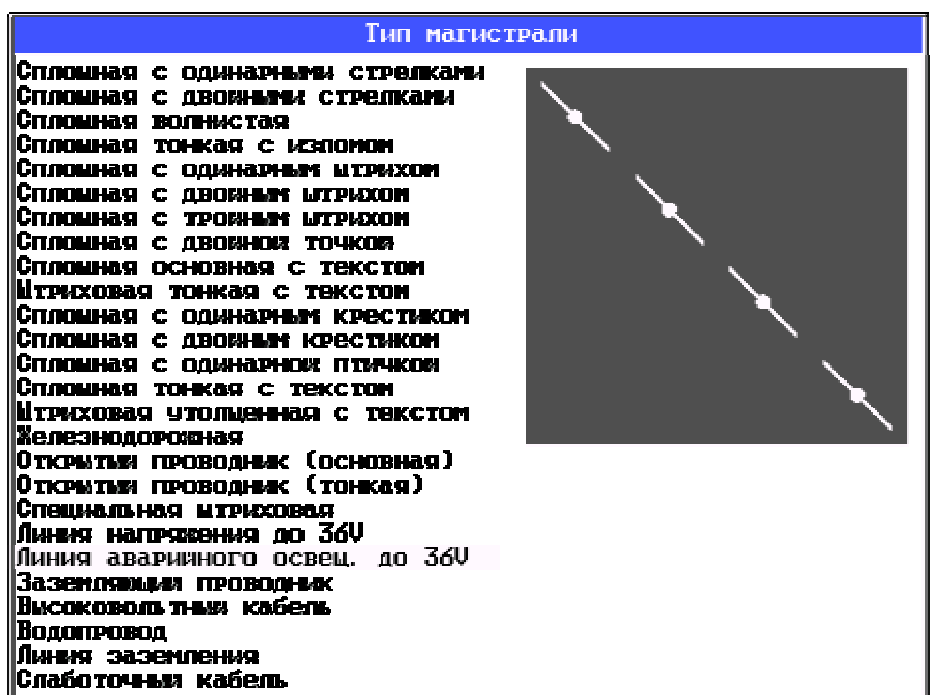

Рис.3. Выбор типа магистрали в меню

Специальный тип геометрического элемента "Магистраль" упрощает вычерчивание условных обозначений трасс кабельных прокладок, проводок, линий трубопроводов и др. в виде линий по ГОСТ 21.106, 21.107, 21.108, 21.614, 2.754. Трудоемкость вычерчивания их вручную несравнима с реализованной, когда задается только осевая линия. Имеется 26 типов магистралей (рис.3). На них остановимся более подробно, как на примере глубокого проникновения стандартов в структуры хранения данных и в программный код САПР. Эти изображения отличаются значительной регулярностью: они состоят из многократно повторяющихся вдоль трассы относительно коротких элементов, что позволяет применять высокоавтоматизированные способы их нанесения, сводящиеся к автоматической генерации условных графических изображений по их параметрическому представлению. За счет параметрического представления компактность хранения магистралей в чертеже повышается радикально.

Параметрическое представление магистралей состоит из двух частей. Главная по объему информации часть помещается в коде графического ядра САПР и идентифицируется байтовым типом магистрали. Остальные параметры размещаются в самих элементах чертежа типа "Магистраль". Это "несущая линия" и "установки". Несущая линия (отрезок прямой или дуга окружности, 21 байт для четырехбайтных вещественных чисел) задает представление магистралей в геометрических преобразованиях: в объектных привязках, при усечении и удлинении других элементов чертежа и самих магистралей, их повороте и растяжении и др. Два типа магистралей в соответствии с их назначением могут включаться в контуры заштрихованных областей: сплошная волнистая и сплошная тонкая с изломами.

Имеются 4 "общие" установки (7 байт): тип магистрали (1 байт) и три установки с диапазоном от 0.01 мм до 600 мм с шагом 0.01 мм, (по 2 байта):
- длина несущей линии в одном шаге (кроме сплошной волнистой, железнодорожной, специальной штриховой, линии аварийного освещ. до 36V, заземляющего проводника, высоковольтного кабеля, водопровода и слаботочного кабеля);
- длина "картинки", помещаемой в просвет несущей линии (кроме линии напряжения до 36V, линии аварийного освещ. до 36V и заземляющего проводника);
- длина несущей линии на первом шаге (кроме сплошной волнистой, железнодорожной и заземляющего проводника).

Есть магистрали без "картинок" – например, железнодорожная, и наоборот, такие, где "картинка" занимает всю длину периода – специальная штриховая, которая позволяет изобразить "зебру" пешеходного перехода. Кроме перечисленных "общих", есть еще 12 байт установок, которые обрабатываются более индивидуально для каждого типа магистрали, как это показано в таблице применимости магистралей различных типов. Если диапазон изменения параметра в таблице не указан – значит, он меняется от 0.1 до 25 мм с шагом 0.1 мм.

Таблица. Объекты, изображаемые магистралями, и их индивидуальные установки

| Объект | Тип магистрали | Индивидуальные установки |
|---|---|---|
| 1 | 2 | 3 |
| Кабели низкого напряжения | Сплошная с одинарными стрелками | продольное отклонение стрелки 0.1 – 12 мм поперечное отклонение стрелки |
| Кабели высокого напряжения | Сплошная с двойными стрелками | то же плюс расстояние между стрелками |
| Линии обрыва, линии разграничения вида и разреза, гибкие трубопроводы, шланги | Сплошная волнистая | длина полуволны от 0.01 до 300 мм высота полуволны от 0.1 до 150 мм |
| Длинные линии обрыва | Сплошная тонкая с изломом | высота излома |
| Канализация хозяйственно-фенольных стоков | Сплошная с одинарным штрихом | высота штриха отклонение штриха от вертикали |
| Канализация загрязненных стоков | Сплошная с двойным штрихом | высота штриха расстояние между штрихами отклонение штриха от вертикали |
| Дополнительная линия, объект по усмотрению | Сплошная с тройным штрихом | высота штриха расстояние между штрихами отклонение штриха от вертикали |
| Дополнительная линия, объект по усмотрению | Сплошная с двойной точкой | расстояние между точками |

| 1 | 2 | 3 |
|---|---|---|
| Проектируемые трубопроводы на чертежах санитарно-технических систем (наружных сетей водоснабжения, канализации, тепловых сетей и т.д.) | Сплошная основная с текстом | размер (от 2.5 до 40 мм) и угол наклона шрифта (75° или 90°) текст от 0 до 4 символов сжатие шрифта 0.1 – 2.55 с шагом 0.01 |
| Невидимые участки существующих трубопроводов на чертежах санитарно-технических систем (сетей водоснабжения, канализации, тепловых сетей и т.д.), а также инженерные сети, прокладываемые в траншее | Штриховая тонкая с текстом | то же |
| Водопровод прямой фильтрованной воды | Сплошная с одинарным крестиком | высота крестика |
| Водопровод обратной фильтрованной воды | Сплошная с двойным крестиком | высота крестика расстояние между крестиками |
| Канализация ливневых стоков | Сплошная с одинарной птичкой | высота птички расстояние между концами птички |
| Существующие трубопроводы на чертежах санитарно-технических систем (наружных сетей водоснабжения, канализации, тепловых сетей) | Сплошная тонкая с текстом | размер (от 2.5 до 40 мм) и угол наклона шрифта (75° или 90°) текст от 0 до 4 символов сжатие шрифта 0.1 – 2.55 с шагом 0.01 |
| Невидимые участки проектируемых трубопроводов на чертежах санитарно-технических систем (сетей водоснабжения, канализации, тепловых сетей и т.д.) | Штриховая утолщенная с текстом | то же |
| Железнодорожные пути | Железнодорожная | расстояние от оси до рельса 0.01 – 600 мм |
| Открытая прокладка проводника сплошной основной линией | Открытый проводник (основная) | высота излома |
| Открытая прокладка проводника сплошной тонкой линией | Открытый проводник (тонкая) | высота излома |
| Различные нестандартные виды штриховых линий | Специальная штриховая | толщина штриха от 0.01 до 600 мм |
| Линии напряжения 36 V и ниже | Линия напряжения до 36V | диаметр точки от 0.01 до 25.5 мм |
| Линии напряжения 36 V и ниже для аварийного освещения | Линия аварийного освещ. до 36V | длина просвета от 0.01 до 600 мм длина штриха от 0.01 до 600 мм диаметр точки от 0.01 до 25.5 мм |
| Заземляющие проводники | Заземляющий проводник | расстояние между штрихами 0.01 – 600 мм длина штриха от 0.01 до 600 мм высота крестика от 0.01 до 25.5 мм |
| Высоковольтные кабели | Высоковольтный кабель | длина просвета от 0.01 до 600 мм продольное откл. стрелки 0 – 12 мм поперечное откл. стрелки 0.05 – 25 мм |
| Водопроводы | Водопровод | длина просвета от 0.01 до 600 мм расстояние между линиями: |
| Заземляющие проводники | Линия заземления | диаметр точки от 0.01 до 25.5 мм |

| 1 | 2 | 3 |
|---|---|---|
| Слаботочные кабели | Слаботочный кабель | длина просвета от 0.01 до 600 мм<br>длина штриха от 0.01 до 600 мм<br>диаметр точки от 0.01 до 25.5 мм |

Все перечисленные сведения об одной магистрали занимают в чертеже 40 байт. К ним, как и к другим элементам чертежа, добавляются номер слоя, цвет, тип элемента чертежа – еще 3 байта. В результате показанные на рис.4 две магистрали занимают в чертеже (на диске и в оперативной памяти) по 43 байта. Для сравнения: отрезок прямой и дуга окружности занимают соответственно 19 и 23 байта. Если бы изображения магистралей рис.4 создавались как отрезки, ломаные, дуги и тексты, то идущая по дуге заняла бы 313 байт, а прямая - 1056 байт.

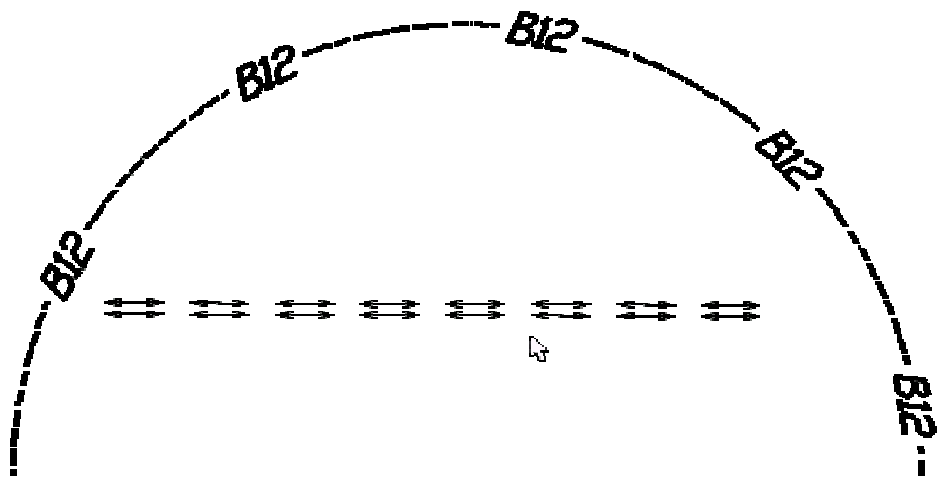

Рис.4. Магистрали в чертеже с несущими линиями в виде дуги и отрезка

Для пользователя процедура построения магистрали выглядит лишь чуть сложнее построения отрезка или дуги. Выбирается нужный тип магистрали из меню (рис.3), выбирается вариант геометрии (отрезок или дуга), производится построение отрезка или дуги с обычными возможностями объектных привязок, ортогонализации, ввода чисел и др., затем перемещением курсора устанавливается положение картинок относительно начала магистрали. Все установки магистрали можно изменить в последующем.

*Заключение*

Изложенные в настоящей работе методы обеспечения требований ЕСКД и СПДС в САПР промышленных объектов не претендуют на полноту, но содержат большое число примеров реализации этой непростой задачи в условиях комплексирования САПР на основе собственного графического ядра. Работоспособность этих методов проверена десятилетним опытом эксплуатации и развития комплексной САПР реконструкции предприятий.

## Литература